# A Finite Element Method for Solving 2D Contact Problems with Coulomb Friction and Bilateral Constraints

Jie Zhang*, Qi Wang#

| | |
|---|---|
| * School of Aeronautic Science and Engineering | # School of Aeronautic Science and Engineering |
| Beihang University | Beihang University |
| BeiHang University XueYuan Road | BeiHang University XueYuan Road |
| No.37,HaiDian District, Beijing, China | No.37,HaiDian District, Beijing, China |
| zhjhn1984@163.com | bhwangqi@sina.com |

**Abstract**

Based on the plenty method, this paper describes a numerical method for 2D non-smooth contact problems with Coulomb friction and bilateral constraints and its application to the simulation of statics and dynamics for a frictional translational joint. Comparison is made with results obtained using a finite element program, ANSYS. Two interesting phenomena in the simulation of the translational joint reveal that hyperstatic problems with Coulomb friction can't be solved just by counting deformation factors and small deformation can affect the rigid motion significantly in some cases.

**1 Introduction**

Contact mechanics studies the behavior of loaded structures in mutual contact. The large amount of researches and efforts devoted to contact problems during the last two decades reveals the importance of the phenomena. Various numerical methods and computational techniques have been proposed for the different classes of the problem, involving different nonlinearities due to the material property, finite geometry changes, or friction effects[1]. The contact problem is inherently a nonlinear problem. The finite element method(FEM) is one of the most efficient tools for solving contact problems with Coulomb friction[2]. There are mainly two methods for modeling and simulation for the normal contact problem in the FEM code: one that is the Penalty method; the other is the Lagrange multiplier methods. Weyler et al. [3] made a comparison of penalty and Lagrange multiplier implementations. Hüeber and Wohlmuth[4] considered residual and equilibrated error indicators for contact problems with Coulomb friction. Maier and Z.Q. Xie et al. [5,6] developed numerical methods to solve the three-dimensional problems with impacts and Coulomb friction. By introducing some appropriate assumptions and analyzing the FEM numerical results, an approximate model for the contact problem of cylindrical joints with clearances is developed through modeling the pin as a rigid wedge and the elastic plate as a simple Winkler elastic foundation[7]. Several FEM strategies and algorithms are presented to solve the unilateral contact problem with Coulomb friction[8-13]. As to the system with bilateral constraints, few study have attempted so far with FEM[14]. Translational joint is used widely in mechanical systems, it's a classical contact problem with bilateral constraints. One difficulty of bilateral constraints contact problems modeling is that it is not easily extendible to contacts involving multiple contact states and multiple contact points. In recent years, extensive work has been done to study the modeling and simulation of non-smooth multi-rigid-body systems with bilateral constraints. Flores et al. [15] used the non-smooth dynamics approach to model the planar rigid multibody systems with translational clearance joints and simulated the dynamic response of a planar slider-crank mechanism with slider clearance. Klepp[16,17] studied the existence and uniqueness of solutions for a single degree of freedom system with two friction-affected translational joints. In his paper, the clearance sizes and the impacts of translational joints were neglected, the translational joints were treated as bilateral constraints. Fangfang Zhuang[18] and Xiaoming Luo et al. presented modeling and simulation methods for the rigid multibody system with frictional translational joints. However, one of the disadvantages of the methods based on rigid model is that they are incapable of hyperstatic problems. Thus research on deformable friction-affected translational joints with Coulomb friction and bilateral constraints is significantly.

Based on the plenty method, this paper describes a numerical algorithm for 2D non-smooth contact problems with Coulomb friction and bilateral constraints and its application to the simulation of statics and dynamics for a frictional translational joint. Comparison is made with results obtained using a finite element program, ANSYS.

The physical significance of the Penalty method is that fictitious springs apply on nodes of the contact surface to simulate the contact forces, base on the Plenty method[3], introducing fictitious linear springs between contact nodes and the constraint surfaces. For the normal direction, the contact model is expressed as a linear function between the normal contact force and normal nodal displacement; for the tangential direction, two cases have to be distinguished, including sticking and slipping. During sticking the tangential contact force is expressed as a linear function of tangential nodal displacement; during slipping, the tangential fictitious springs are eliminated and the tangential contact force is expressed as a linear function of the normal contact force of the node. During sticking, the problem is treated as a static problem and is solved by static analysis using FEM. Based on the kineto-elastodynamic(KED) method, a reliability analysis method for 2D dynamic contact problems with Coulomb friction is put forward in this paper which can avoid the oscillations in the accelerations. The method can be generally used and easily implemented in computer programs.

The clearance sizes of the translational joints are so small that the impacts between the sliders and the guides can be neglected, therefore the geometric constraints of the translational joints are treated as bilateral constraints. One difficulty of contact problems with bilateral constraints is that it involves multiple contact modes and multiple contact points. A trial-and-error and iterative method is capable of solving this problem.

There are two interesting phenomena in the simulating of the translational joints: one is a special hyperstatic problem which is different from the general hyperstatic problems, it can't be solved by counting deformation factor only; the other is the dynamics of the translational joints which reveal that small deformation affects the Rigid motion significantly.

## 2. Model for flexible system with Coulomb friction

Sections 2.1 and 2.2 describe a numerical algorithm for 2D non-smooth contact problems with Coulomb friction and its application to the simulation of sticking and slipping with finite element method. These sections cover Coulomb's friction model, Penalty method and the method proposed in this paper for simulating stick and slip behaviors based on the Penalty method. The method can be used in general static and dynamic contact problems with Coulomb friction and deformation.

## 2.1. Coulomb friction model

Friction is a natural occurrence that affects almost all objects in motion. Coulomb's law is frequently used to describe the friction phenomenon for contact problems. It represents the most fundamental and simplest model of friction between dry contacting surfaces. The main problem with Coulomb friction is the discontinuity of the friction force due to the difference between static and dynamic behaviors. Although Coulomb friction is the classical one, it causes difficulties of convergence and its mathematical treatment remained open for a long time. The main difficulty results from instantaneous changes in the contact forces at transitions from sliding to sticking or transitions from sticking to sliding. Different models have been developed to permit a smooth transition from sticking to sliding friction resulting in distortion.

Some work has been done in previous studies based on Coulomb friction. The Coulomb's friction law states that the friction force acts tangent to the contacting surfaces in a direction opposed to the relative motion or tendency for motion of one surface against another and the magnitude of friction force $F_T$ is proportional to the magnitude of the normal contact force $F_N$, at the contact point by introducing a coefficient of friction $\mu$. When the relative tangential velocity of the contacting bodies is not zero, the Coulomb's friction law is given by

$$F_T = -\mu F_N \operatorname{sgn}(v_T), \qquad v_T \neq 0 \tag{1}$$

where $v_T$ is relative tangential velocity, $\mu$ is the coefficient of kinetic friction. When the relative tangential velocity is zero, the friction force has a value within a range given by

$$-\mu' F_N \leq F_T \leq \mu' F_N, \qquad v_T = 0 \tag{2}$$

where $\mu'$ is the coefficient of static friction. $\mu$ is generally smaller than $\mu'$.

When the relative tangential velocity is zero, the friction force is dependent on acceleration. The acceleration and friction force are coupling, which becomes a difficult point for simulation, it's expressed as

$$F_T = -\mu' F_N \operatorname{Sgn}(a_T) \qquad v_T = 0 \tag{3}$$

$\operatorname{Sgn}(x)$ is multifunction, which is defined as

$$\operatorname{Sgn}(x) := \begin{cases} +1 & x > 0 \\ [-1, +1] & x = 0 \\ -1 & x < 0 \end{cases} \tag{4}$$

Two cases have to be distinguished: sticking and slipping. When both of the relative tangential velocity and acceleration both are zero, corresponding to no motion and no tendency for motion, this phenomenon is defined as sticking; otherwise defined as slipping. In the case of slipping, Coulomb friction model states that the friction force is proportional to the magnitude of the normal contact force. If the relative velocity and acceleration vanish, sticking takes place and the frictional force is equal to the external forces in the tangential direction.

The Coulomb's friction model is illustrated in Fig.1.

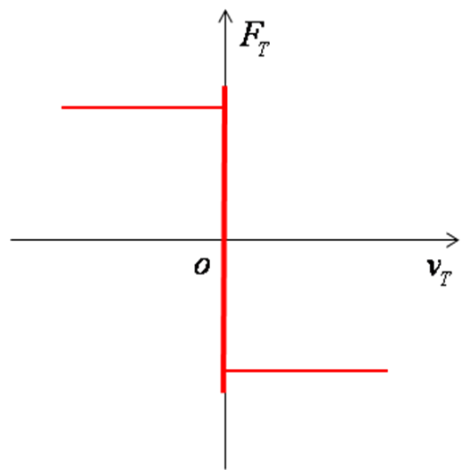

**Figure 1.** The Coulomb friction model

**2.2    A FEM method for flexible system with Coulomb friction**

**2.2.1   Preliminary assumptions based on Penalty method**

There are mainly two methods in modeling and simulating for the normal contact problem in the FEM code, including Penalty method and Lagrange multiplier method.

In the Penalty method, the contact forces are proportional to the quantity of penetration by introducing a plenty number which is physically equivalent to an additional fictitious linear spring between contacted bodies. The finite element implementation of the Penalty method is discussed in detail in textbook.

Based on the Penalty methods, this paper describes a numerical algorithm for 2D non-smooth contact problems with Coulomb friction which is applicable to both static and dynamic conditions.

It is assumed that contact forces are developed between contact nodes and target surfaces in FEM, as illustrated in Fig.2. As mentioned above, the physical significance of the Penalty method is applying fictitious springs on nodes of the contact surface to simulate the contact forces. Based on the Plenty method, the model for contact problems with Coulomb friction is developed. Similar to Winkler elastic foundation theories with normal fictitious springs used on the surface[7], surfaces of constraint are modeled by adding plenty of fictitious linear springs in the normal and tangential directions in the new model. These fictitious springs, independent to each other, are added between contact nodes and contact surfaces, as illustrated in Fig.3 and 4, which simulate not only the elasticity of the constraint surfaces, but also the interaction between the contact body and the constraint surfaces. These springs are fictitious which has numerical meaning only, although they are pictured out in Figures.

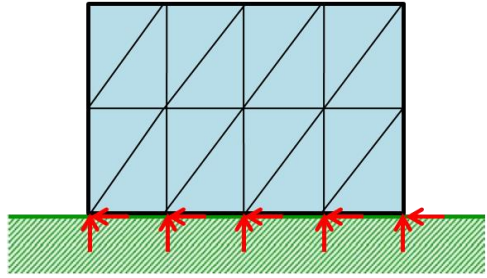

**Figure2.** Diagram of the nodal contact forces in FEM

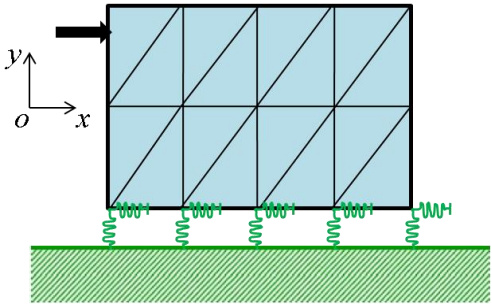 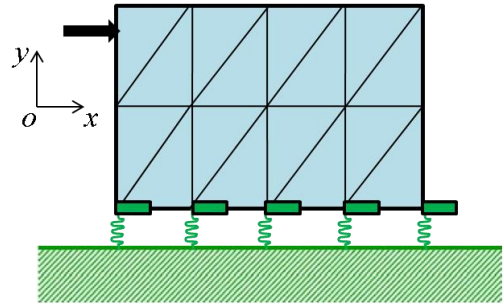

**Figure 3.** Diagram for modeling of slipping condition

**Figure 4.** Diagram for modeling of slipping condition

For the normal direction, the contact model is expressed as a linear function between the normal contact force and normal nodal displacement; for the tangential direction, the tangential contact force is expressed as a linear function of tangential nodal displacement during sticking solved by static analysis using FEM, as illustrated in Fig.3; during slipping, the tangential fictitious springs are eliminated and the tangential friction force is expressed as an linear function of the normal contact force of the contact nodes, as shown in Fig.4, the green rectangles indicate the linear relationship between tangential friction force and normal contact forces at contact nodes. The stiffness of the fictitious springs depends on material and geometric properties et cl, and can be obtained by calculate theory in Penalty method. The model is convenient for programming.

### 2.2.2 Judgment between sticking and slipping

In the FEM, interaction occurs at all of the nodes. Generally speaking, there are a large amount of nodes in the system, a lot of work of calculation is required if determination of the state (sticking or slipping) is performed at each node, only with limited advantages. With the expected accuracy being obtained, the determination of the state focuses on the whole system (i.e. all of the nodes are under the same state), in this paper. Such an assumption is made to simplify the mathematical analysis with agreement of practice and significant reduction of work of calculation. Since the change of velocity is continuous without any sudden change, the sum of the tangential relative velocity of the contact nodes at the last moment is selected for the preliminary criterion of the state.

This paper treats sticking as a static problem and solves it by static analysis using FEM, since it refers to static problems essentially.

The sum of the tangential relative velocity of the contact nodes is described as $V_T$.

Firstly, the sum of the tangential relative velocity $V_T$ is calculated by

$$V_T = \sum_{i=1}^{n} V_{Ti} \tag{5}$$

where $i$ is the number of contact nodes, $V_{Ti}$ is the velocities of contact nodes at the previous time step.

If $V_T \neq 0$, it is obviously that the contact state is slipping.

If $V_T = 0$, a trial-and-error method is applied for the state judgment between sticking and slipping.

The static analysis using FEM is implemented and the trial friction forces $F_{Ti}$ and the trial normal forces $F_{Ni}$ of each contact node are obtained.

Maximum static friction is calculated for the judgment. In Coulomb friction model, the criterion for the direction of the maximum static friction force is depending on acceleration, which is expressed as

$$F_T = -\mu' F_N \operatorname{Sgn}(a_T) \tag{6}$$

The coupling between friction force and acceleration is a difficult point for simulation, in this paper the criterion for the direction of friction force is varied from acceleration to the trial deformation which makes simulation more convenient. The equation is written as

$$F_{Ti} = -\mu' F_{Ni} \operatorname{sgn}(\Delta q) \quad v_T = 0 \text{ and } a_T \neq 0 \tag{7}$$

where $\Delta q_T$ is the sum of the tangential displacement of the contact nodes which is calculated by

$$\Delta q_T = \sum_{i=1}^{n} \Delta q_{xi} \tag{8}$$

where $i$ is the number of contact nodes. $\Delta q_{xi}$ is the nodal tangential displacement of the contact nodes.

Then the sum of the limiting value of friction (maximum static friction) $F_T^{\max}$ can be calculated by

$$F_T^{\max} = \sum_{i=1}^{n} \mu' |F_{Ni}| \operatorname{sgn}(\Delta q_T) \tag{9}$$

Sticking takes place if the sum of the trial tangential friction forces of contact nodes doesn't exceed the sum of the maximum static friction force $F_T^{\max}$. Once the sum of the maximum static friction force $F_T^{\max}$ has been reached, sliding takes place.

The schematic procedure is briefly shown as follows.

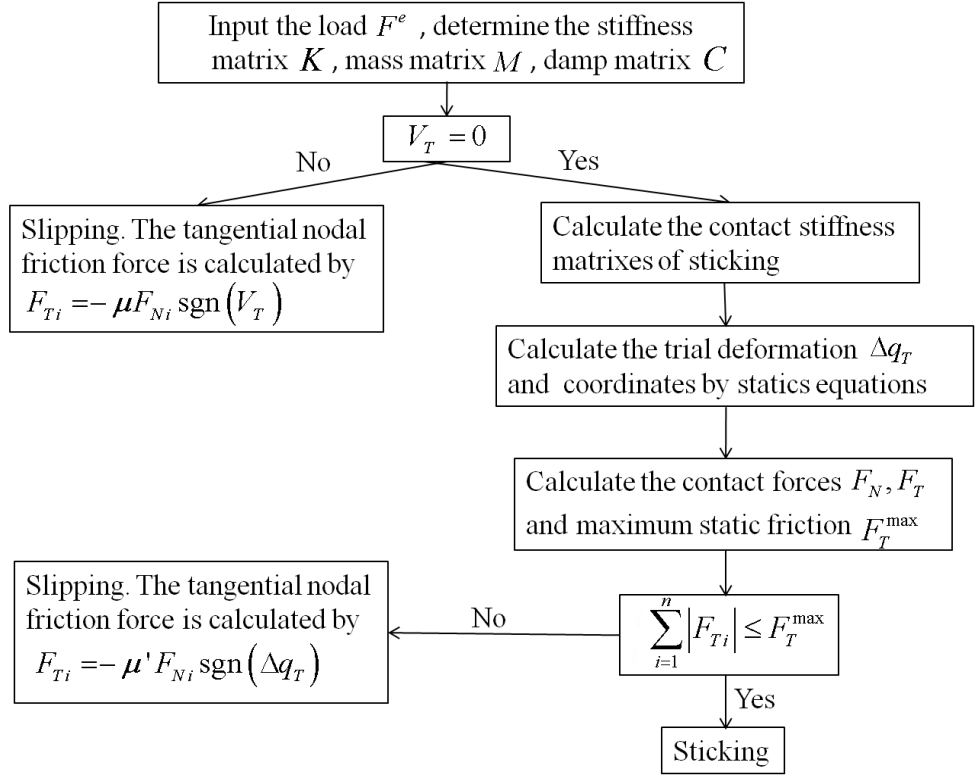

**Figure 5.** The schematic procedure of the judgment between sticking and slipping

### 2.2.3 Modeling of sticking condition

Most commercial FEM software simulates sticking condition by dynamic analysis. In these treatments a crucial issue is arising, oscillations in the acceleration. Actually, during sticking there is no motion and no tendency toward motion, and it belongs to a static problem, according to the mechanical characteristic of sticking, this paper treats the system as a static problem and solves it by a static analysis using FEM which is different from other treatments.

Although there is no motion and no tendency for motion, deformation caused by the action of tangential forces is further complicated by micro-slip within the contact region. In this paper, the micro-slip is simulated as tangential displacements of contact nodes, these nodes are also the acting points of friction forces, the tangential friction forces are simulated as forces of the tangential fictitious springs which are added between the contact nodes and the contact surfaces. If tangential forces exceed the limiting value of friction (maximum static friction), contact nodes will slide completely. Otherwise, there may be no sliding and just micro-slip.

According to the Coulomb's friction law described in section 2.1, during sticking, the friction force has a value within a range given by

$$-\mu' F_N \leq F_T \leq \mu' F_N \quad v_T = 0 \text{ and } a_T = 0 \tag{10}$$

The inequality constraints are caused by Coulomb friction. The friction forces can also be expressed as the other way

$$F_T = -F_T^e \quad v_T = 0 \text{ and } -\mu' F_N \leq F_T^e \leq \mu' F_N \tag{11}$$

where $F_T^e$ is a sum of the tangential external forces, and the criterion of sticking has been transited from $v_T = 0 \, and \, a_T = 0$ into $-\mu' F_N \leq F_T^e \leq \mu' F_N$. In general cases, $F_T^e$ is an unknown quantity due to it is always has a relationship with the normal contact force $F_N$, so the formula(11) is rarely used as a criterion for judging sticking and slipping, especially based on the rigid body model. Based on the flexible model and numerical approach presented in this paper, formula(11) is used as the criterion of sticking and avoids the inequality constraints which is illustrated in formula (10).

A model with plenty of numerical fictitious linear springs between contact nodes and the constrain surfaces has been developed, as illustrated in Fig.3, 4. The interaction between elastic contacting bodies is transmitted via plenty of fictitious linear springs at contact nodes. According to the static equilibrium principle of force, the tangential friction forces equals to the tangential external loads.

The model for contact forces of each node is described as follows, $i$ is the number of one contact node. The contact force model is expressed as a linear function between the normal contact force $F_{Ni}$ and normal nodal displacement for the normal direction

$$F_{Ni} = -k_N q_{yi} \tag{12}$$

where $q_{yi}$ is the normal nodal displacement of the contact nodes, $k_N$ is the normal fictitious spring stiffness which can be determined by calculation theory in penalty method.

In the tangential direction, the tangential contact force $F_{Ti}$ is expressed as a linear function of tangential nodal displacement which is expressed by equation

$$F_{Ti} = -k_T q_{xi} \tag{13}$$

where $q_{xi}$ is the tangential nodal displacement of the contact nodes, $k_T$ is the tangential fictitious spring stiffness which can be determined by calculation theory in penalty method[3].

The contact force vector can be expressed by equation:

$$F_N = K_N q \tag{14}$$

$$F_T = K_T q \tag{15}$$

where $F_N$ is the normal contact force vector, $F_T$ is the tangential contact force vector, $q$ is the displacement vector of the system, $K_N$ is the normal contact stiffness matrix, $K_T$ is the tangential contact stiffness matrix. These stiffness matrixes are expressed as follows, where $l$ is the number of the nodes which come into contact with the surfaces.

The normal stiffness matrix $K_N$ is expressed as

$$K_N[2l+1][2l+1] = -k_N \tag{16}$$

else
$$K_N[m][n] = 0 \quad m = n \neq 2l+1 \quad or \quad m \neq n \tag{17}$$

The tangential stiffness matrix $K_T$ is expressed as

$$K_T[2l][2l] = -k_T \tag{18}$$

else
$$K_T[m][n] = 0 \quad m = n \neq 2l \quad or \quad m \neq n \tag{19}$$

Based on the equations of the contact force model, the static equations can be expressed as:

$$Kq = F^e + F_N + F_T \tag{20}$$

substitute (14) and (15) into (20) and results in

$$Kq = F^e + K_N q + K_T q \tag{21}$$

Revises the stiffness matrix by the shift of Items and the equation can be rewritten as

$$(K - K_N - K_T)q = F^e \tag{22}$$

Thus, base on the plenty method, the equation of planar motion for the deformable body with coulomb friction can be formulated in the form of a linear algebraic system.

$$K'q = F^e \tag{23}$$

where $K'$ is the revised stiffness matrix which is calculated by

$$K' = K - K_N - K_T \tag{24}$$

The numerical approach for linear algebraic equations is a textbook matter. It can be calculated conveniently.

In some cases with complex constraints, the determination of the contact region must be determined before calculation. The application of a trial-and-error method can solve this problem. Which are discussed in detail in section 3.2.

The advantage of the method mentioned above is that it reveals mechanical mechanism of sticking and avoids the inequality Constraints caused by Coulomb friction, it's convenient for calculation.

### 2.2.4 Modeling of slipping condition

Once the sum of tangential friction forces exceeds the maximum static friction force, the condition of system changes from sticking to sliding instantaneous. As mentioned above, In the case of slipping, there is motion or tendency for motion, and it belongs to dynamics.

According to the Coulomb's friction law which is mentioned in section 2.1, the magnitude of nodal friction force is proportional to the magnitude of the normal contact force at each node during slipping. A deformable model for 2D contact problems during slipping is put forward in this paper. At each node, for the normal direction, the contact model is expressed as a linear function between the normal contact force and normal nodal displacement; for the tangential direction, the tangential fictitious springs are eliminated and the nodal friction force is expressed as a linear function of the normal contact force of the contact nodes.

The analysis of models for contact forces will be discussed in two cases during slipping: In case 1, the relative tangential velocity of contact body is not zero; in case 2, the relative tangential velocity of contact body is zero and the tangential acceleration is zero.

In case 1, the model for contact forces of each node is described as follows, $i$ is the number of one contact node. The contact force model is expressed as a linear function between the normal contact force $F_{Ni}$ and normal nodal displacement for the normal direction:

$$F_{Ni} = -k_N q_{yi} \quad (25)$$

where $q_{yi}$ is the normal nodal displacement of the contact nodes, $k_N$ is the normal fictitious spring stiffness which can be determined by calculation theory in penalty method.

In the tangential direction, the tangential contact force $F_{Ti}$ is expressed as a linear function of normal nodal force $F_{Ni}$ according to the Coulomb's friction law

$$F_{Ti} = -\mu F_{Ni} \operatorname{sgn}(v_T) \quad (26)$$

where $v_T$ is relative tangential velocity described in section 2.2.2.

substitute (25) into (26) and obtain:

$$F_{Ti} = -\mu k_N q_{yi} \operatorname{sgn}(v_T) \quad v_T \neq 0 \quad (27)$$

The contact force vector can be expressed by equation:

$$F_N = K_N q \quad (28)$$

$$F_T = K_T q \quad (29)$$

where $F_N$ is the normal contact force vector, $F_T$ is the tangential contact force vector, $q$ is the displacement vector of the system, $K_N$ is the normal contact stiffness matrix, $K_T$ is the tangential contact stiffness matrix. These stiffness matrixes are expressed as follows, where $l$ is the number of the nodes which come into contact with the surfaces.

The normal stiffness matrix $K_N$ is written as

$$K_N[2l+1][2l+1] = -k_N \quad (30)$$

else $\quad K_N[m][n] = 0 \quad m = n \neq 2m+1 \text{ or } m \neq n \quad (31)$

The tangential stiffness matrix $K_T$ is described as

$$K_T[2l][2l+1] = -\mu k_N \operatorname{sgn}(v_T) \tag{32}$$

else
$$K_T[m][n] = 0 \quad m \neq 2l \text{ or } n \neq 2l+1 \tag{33}$$

In case 2, the model for contact forces of each node is described as follows, $i$ is the number of one contact node, the contact force model is expressed as a linear function between the normal contact force $F_{Ni}$ and normal nodal displacement for the normal direction as the same as case 1.

$$F_{Ni} = -k_N q_{yi} \tag{34}$$

where $q_{yi}$ is the normal nodal displacement of the contact nodes, $k_N$ is the normal fictitious spring stiffness which can be determined by calculation theory in penalty method.

In the tangential direction, the tangential contact force $F_{Ti}$ is expressed as a linear function of normal contact force $F_{Ni}$. As described in section 2.2.2, the criterion for the direction of friction force is varied from acceleration to the trial displacement which is different from Coulomb friction model and become more convenient for simulation. The equation is written as

$$F_{Ti} = -\mu' F_{Ni} \operatorname{sgn}(\Delta q_T) \quad v_T = 0 \text{ and } a_T \neq 0 \tag{35}$$

where $\Delta q_T$ is the trial displacement which is calculated by equation(8)

substitute (34) into (35) and obtain:

$$F_{Ti} = -\mu' k_N q_{yi} \operatorname{sgn}(\Delta q_T) \quad v_T = 0 \text{ and } a_T \neq 0 \tag{36}$$

The contact force vector can be expressed by equation:

$$F_N = K_N q \tag{37}$$

$$F_T = K_T q \tag{38}$$

where $F_N$ is the normal contact force vector, $F_T$ is the tangential contact force vector, $q$ is the displacement vector of the system, $K_N$ is the normal contact stiffness matrix, $K_T$ is the tangential contact stiffness matrix. These stiffness matrixes are expressed as follows, where $l$ is the number of the nodes which come into contact with the surfaces.

The normal stiffness matrix $K_N$ is written as

$$K_N[2l+1][2l+1] = -k_N \tag{39}$$

else $\quad K_N[m][n]=0 \quad m=n \neq 2m+1 \ or \ m \neq n$ (40)

The tangential stiffness matrix $K_T$ is described as

$$K_T[2l][2l+1]=-\mu k_N \, \text{sgn}(\Delta q_T) \tag{41}$$

else $\quad K_T[m][n]=0 \quad m \neq 2l \ or \ n \neq 2l+1$ (42)

Based on the equations of the contact force model mentioned above, the dynamic equations in two cases can be expressed as:

$$M\ddot{q}+C\dot{q}+Kq=F^e+F_N+F_T \tag{43}$$

where $M$ is mass matrix, $C$ is damp matrix, $K$ is the stiffness matrix, $F^e$ is the equivalent external nodal force vector, which are calculated by classical procedure of FEM. And $\ddot{q}$ is the acceleration vector, $\dot{q}$ is the velocity vector.

Substitute (14) (15) (37) and (38) into (43) and results in

$$M\ddot{q}+C\dot{q}+Kq=F^e+K_N q+K_T q \tag{44}$$

Revises the stiffness matrix by the shift of Items and the equation can be rewritten as

$$M\ddot{q}+C\dot{q}+(K-K_N-K_T)q=F^e \tag{45}$$

Thus, base on the plenty method, the equation of planar motion for the deformable body with Coulomb friction can be formulated in the form of a linear algebraic system.

$$M\ddot{q}+C\dot{q}+K'q=F^e \tag{46}$$

where $K'$ is the revised stiffness matrix which is calculated by

$$K'=K-K_N-K_T \tag{47}$$

In time integration methods for dynamic analysis using FEM, such as Newmark method and central difference method, a crucial issue is arising oscillations in accelerations. In the general case, convergence to steady state solutions can be obtained after some time steps. However, in the 2D dynamic contact problems with Coulomb friction and bilateral constraints, there are time-varying contact modes and instantaneous changes in the contact forces at transitions between sliding and sticking. These nonlinearities cause convergence difficulties during slipping. In the initial time steps, convergence to steady state solutions of accelerations is impossible, owing to the contact forces are related to accelerations which oscillate during simulation.

The kineto-elastodynamics(KED) modeling approach has been widely used to model flexible mechanisms [20]. The procedure of kineto-elastodynamics(KED) method is briefly given out. At first, rigid body motion and acceleration can be approximated with rigid body model. Then according to D'Alembert's Principle, this paper performs static analysis using FEM by utilizing inertial forces due to the rigid body motion as external loads for flexible system simulation. KED method provides a convenient vehicle for avoiding the oscillations in the acceleration. The assumptions made here include [21]: (1) rigid body motion due to its coupling with flexible motion is much smaller compared with the rigid body motion due to rigid body driving forces, hence the influence of flexible motion on rigid body motion is negligible; (2) joint constraint forces in the flexible linkage case are close to those in the rigid linkage case, which has been demonstrated to be valid for both open-loop and closed loop structures.

However, in the 2D dynamic contact systems with Coulomb friction and bilateral constraints, the influence of small deformation on rigid body motion is significant, and joint constraint forces in the flexible linkage case are different from those in the rigid linkage case, which are discussed in detail in section 3.3.

Based on the kineto-elastodynamic(KED) method, a simulation method for 2D dynamic contact problems with Coulomb friction is put forward in this paper which can avoid the oscillations in the accelerations and take into account the influence of flexible motion on rigid body motion, furthermore , joint is modeled as a flexible body in the method. The method can be generally used and easily implemented in computer programs. It is described as follows:

Firstly, rigid body motion and acceleration can be approximated with rigid body model[18]. Secondly, according to D'Alembert's Principle, the static analysis using FEM is implemented by utilizing inertial forces due to the rigid body motion as external loads. Thirdly, with the contact forces which are simulated before, a new acceleration can be approximated according to the second law of Newton, and the static analysis using FEM is implemented by utilizing inertial forces which is calculated by the new acceleration. Repeat the cycle continuously until the desired accuracy is obtained. Different from the KED method, the method in this paper take into account the influence of flexible motion on rigid body motion by cycle. Block diagram is shown in Fig.6.

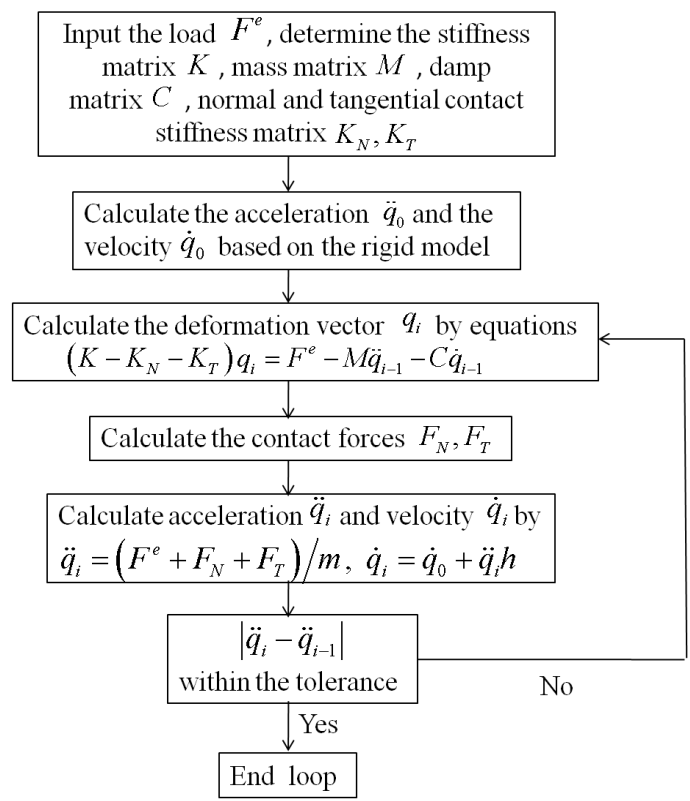

**Figure 6.** Block diagram representation on dynamic simulation of slipping

The method for simulating slip behavior is put forward in this section which can avoid the oscillations in the accelerations and take into account the influence of flexible motion on rigid body motion, furthermore, joint can be modeled as a flexible body in this method. The method can be generally used and easily implemented in computer programs.

## 3 Modeling of translational joint with Coulomb friction

The clearance sizes of translational joint are so small that the impacts between the slider and the guides are neglectable, the geometric constraints of the translational joint are treated as bilateral constraints. One difficulty of bilateral constraints contact problems is that involving multiple contact modes and multiple contact points and brings certain difficulties for numerical simulation.

First of all, the mechanical characteristics of translational joint with Coulomb friction is introduced, then the method for treating sticking and slipping is proposed, this paper reveals two interesting phenomena in the modeling and simulating of the translational joints: one is a special hyperstatic problem which is different from general hyperstatic problems, it can't be solved by counting deformation factor only; the other one is that in the dynamics of translational joint, small deformation affects the rigid body motion significantly during slipping and constraint forces in the flexible linkage case are different from those in the rigid linkage case, which are different from previous research.

The analysis has ignored the normal and tangential damping components of impact and the energy loss to elastic stress waves during the elastic deformation part of impact. This modeling is possible only for low impact speeds. The impacts between sliders and guides are neglected as a consequence of the clearance sizes of the translational joints are very small compared with other linear dimensions.

### 3.1 The mechanical characteristics of translational joint

#### 3.1.1 Contact configurations of translational joint

Figure 7 shows a planar translational joint with clearance. The clearance sizes of the translational joint are so small that the impacts between the slider and the guides are neglectable; the geometric constraints of the translational joint are treated as bilateral constraints. One difficulty of bilateral constraints contact problems modeling is that involving multiple contact modes and multiple contact points.

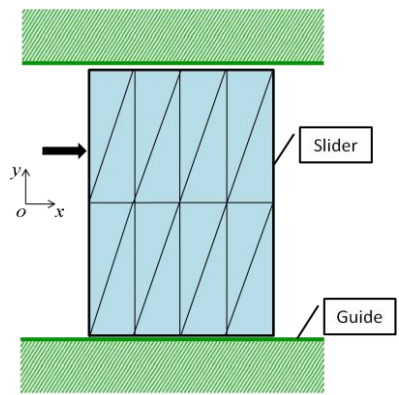

**Figure 7.** Planar translational joint with clearance constituted by a slider and its guides

When the slider moves inside its guide, there are several possible configurations for the relative position between the slider and guide, qualitative analysis of the configurations based on the deformable model is the same as the configurations based on the rigid body model[18], the configurations are illustrated in Figs. 8, 9. The springs in Figs.8, 9 are fictitious which has numerical meaning only, although they are pictured out in Figures. The nodes come into touch with the constraint surfaces is illustrated as green springs, otherwise are the black springs. These different configurations are the following:

Configuration 1: One surface of the slider is in contact with the guide surface, normal contact forces only act on one surface, as shown in Fig.8. It can be treated as unilateral constraint.

Configuration 2: Two opposite slider surfaces are in contact with the guide surfaces as shown in Fig.9, normal contact forces act on two surfaces. It is treated as bilateral constraints.

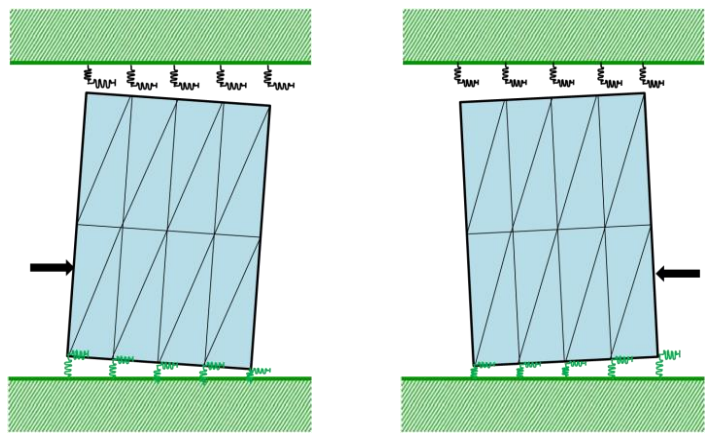

**Figure 8.** One surface of the slider come into contact with its guides

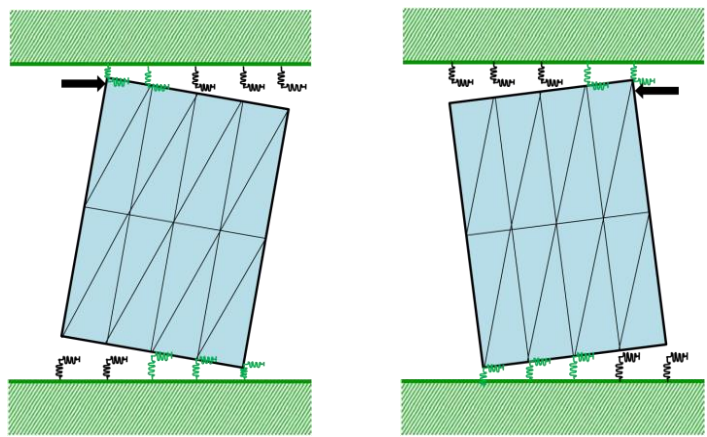

**Figure 9.** Two surfaces of the slider come into contact with its guides

The difficulty results from instantaneous changes in the configurations of contact. There are time-varying contact configurations between the guides and the sliders. One characteristic of translational joint is the analysis of the configurations of contact, an essential component of simulation for system with bilateral constraints. A trial-and-error and iterative method is applied for solving this problem which discusses in detail in section 3.2.

### 3.1.2 Constraints on normal contact forces

As to the normal contact forces, for any configuration of the relative position between the slider and its guide, the magnitudes of the normal contact forces are subjected to the conditions: $F_{Ni}^{U} < 0$ and $F_{Ni}^{L} > 0$, where $i$ is the number of contact nodes, $F_{Ni}^{U}$ is the normal nodal force which is on the upper surface, $F_{Ni}^{L}$ is the normal contact force of contact nodes which is on the lower surface. Owning to the infinitesimal clearance of the translational joint, the normal contact forces of opposite nodes are subjected to the complementarity conditions, $F_{Ni}^{U} \cdot F_{Nk}^{L} = 0$, as illustrated in Fig.9. It is difficult to calculate, accounting for the fact that the system involves plenty of nodes on the surface.

### 3.1.3 Constranits on static friction forces

During sticking, there is no motion and no tendency for motion, and just micro-slip arises which is caused by deformation. In contact problems with unilateral constraint, Once the maximum static friction force has been reached, the system transitions to sliding. These situations can be simulated with the method described in section2.2. But in contact problems with bilateral constraints, crucial issues are arising because of the different states of static frictions of the contacting nodes.

As described in section 2.1, during sticking, the friction force has a value within a range given by

$$-\mu' F_N \leq F_T \leq \mu' F_N \quad v_T = 0 \text{ and } a_T = 0 \tag{48}$$

The analysis on the states of nodal static friction is divided mainly in two conditions. As described as follows

$$F_{Ti} < \mu' F_{Ni} \cdot \text{sgn}(\Delta q_T) \tag{49}$$

$$F_{Ti} = \mu' F_{Ni} \cdot \text{sgn}(\Delta q_T) \tag{50}$$

where $i$ is the number of contact nodes, $F_{Ti}$ is the tangential friction force of the contact nodes, $F_{Ni}$ is the normal contact force of the contact nodes, $\Delta q_T$ is the trial deformation.

In condition A, the nodal friction force is expressed as a linear function of tangential nodal displacement, and a fictitious linear spring is added at the node to model the action; however in condition B, the tangential fictitious spring is eliminated and the magnitude of nodal friction force is proportional to the magnitude of the normal contact force. These two conditions is modeled and simulated in different ways as described in section 2.2. In the contact problem with Coulomb friction and bilateral constraints, when two opposite slider surfaces are in contact with the guide surfaces, these two conditions occur at the same time.

As mentioned above, the characteristics of translational joints cause difficulties in simulation and a new method is required.

## 3.2 Modeling of sticking for translational joint

During sticking, the system with unilateral constraint can be simulated by applying the method expressed in section 2.2.4, however, as to the system with bilateral constraints, the simulation become a difficult point. In the contact system with Coulomb friction and bilateral constraints, when two opposite slider surfaces are in contact with the guide surfaces, not only multiple conditions of static friction occur simultaneously which is described in section 3.1, but also other constraints are imposed which should be obeyed at each node. Except the elastic constitutive models and deformation compatibility equation, there are six constraints which are illustrated as follows:

Equilibrium principle of moment:

$$\sum_{i=1}^{n} M_i = 0 \tag{51}$$

where $i$ denoted as the number of forces, $M_i$ is the moment of forces.

Equilibrium principle of tangential force:

$$\sum_{i=1}^{n} F_{xi} = 0 \tag{52}$$

where $i$ denoted as the number of forces, $F_{xi}$ is the tangential force component.

Equilibrium principle of normal force:

$$\sum_{i=1}^{n} F_{yi} = 0 \tag{53}$$

where $i$ denoted as the number of forces, $F_{yi}$ is the normal force component.

The inequality constraints of normal contact forces:

$$F_{Ni}^{U} < 0 \quad \text{and} \quad F_{Ni}^{L} > 0 \tag{54}$$

The complementarity condition of normal forces of opposite contact nodes:

$$F_{Ni}^{U} \cdot F_{Nk}^{L} = 0 \tag{55}$$

The inequality constraints of the contact forces caused by Coulomb friction:

$$F_{Ti} < \mu' F_{Ni} ; \text{sgn}(\Delta x) \quad \text{or} \quad F_{Ti} = \mu' F_{Ni} \cdot \text{sgn}(\Delta x) \tag{56}$$

where $i$ denoted as the number of contact nodes.

They are physical unequivocal states and can be interpreted as unique solutions for the kinetic problem with specified configuration of the system.

As is known to all, mechanical model is described by the corresponding mathematical model and solution of the mathematical model reveals the validity and reliability of the mechanical model.

A simple numerical example for sticking involving four different conditions indicates that all the constraints mentioned above can not be ignored. The planer slider of mass $m$, length $a$, and width $b$, which is subjected to the action of the force $F^e$, moves in the guide for which the coefficient of static friction is $\mu' = 0.6$. In the analysis 2 linear triangular elements, 4 nodes, and 8 degrees-of-freedom are used as illustrated in Fig 4. The slider is made of an elastic isotropic, homogeneous material characterized by Young's modulus $E = 2.1 \times 10^{11}$ Pa and Poisson's ratio $o = 0.25$ (steel). As shown in Fig.10. The stiffness of the normal and tangential fictitious spring is $K_N = K_T = 2.1 \times 10^{11}$ Pa. Nodal contact forces and external forces are illustrated in the Figs.12, 13, 14, 15. They are calculated by FEM program with different constraints. In these Figures, the green rectangles indicate the linear relationship between tangential friction force and normal contact forces at contact nodes, the nodal contact forces which are in the correct directions are pictured as red arrows, the nodal contact forces which are in the false directions are pictured as purple arrows. External forces and gravity force are pictured as black arrows. The unit of forces and moments illustrated in the Figures are $N$ and $N \cdot m$ respectively. The resultant moment and resultant forces of the system are illustrated in the lower part of the Figures.

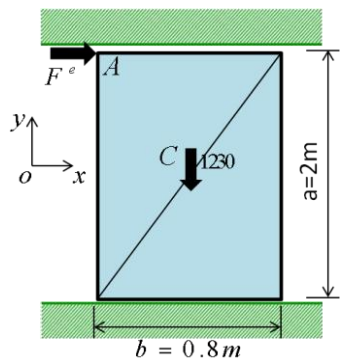

**Figure 10.** Diagram of the slider

In condition 1, a typical numerical approach for general hyperstatic problems is applied to solve the hyperstatic problem during sticking, in which no inequality constraints are imposed, as illustrated in Fig.11. Equations are formulated by the elastic constitutive equation, geometrical compatibility condition, and equilibrium principle of forces and moments as described in equations (51) (52) and (53). The solution of condition 1 illustrates that the mechanical model is improper and the hypestatic problem with Coulomb friction is different from general hypestatic problems and can't be solved by typical approach.

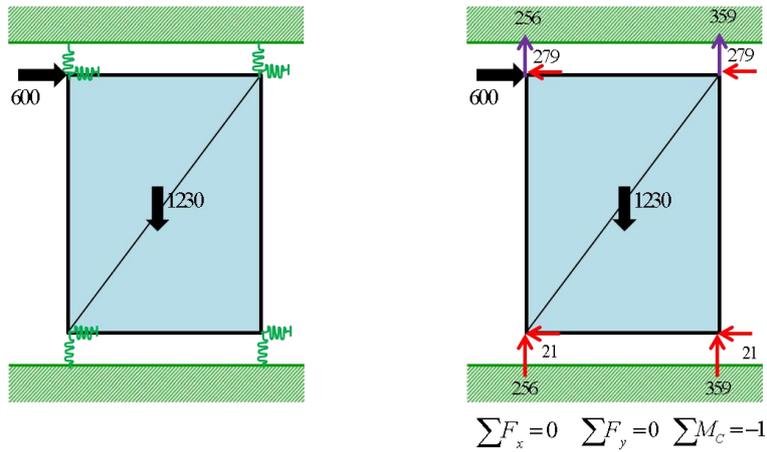

**Figure 11.** Diagram of the contact model and simulation in condition 1

In condition 2, with the consideration of contact model in which two opposite slider surfaces come into contact with the guide surfaces as illustrated in Fig.12. Omission of the inequality constraints as described in equations (54) and (56), the simulation result reveals that the model is improper.

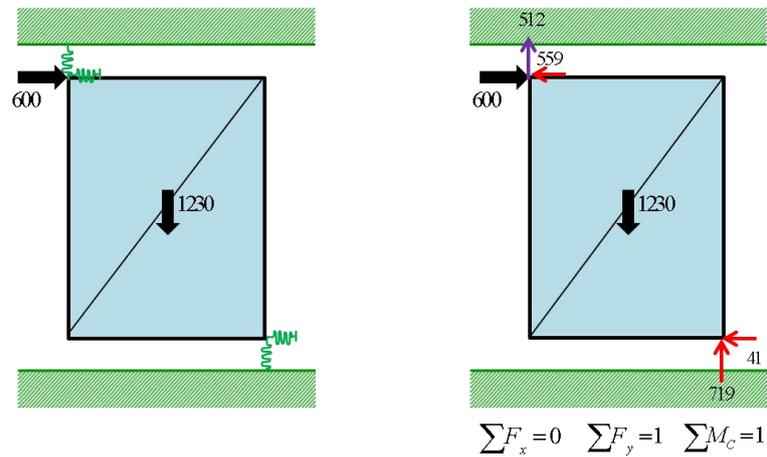

**Figure 12.** Diagram of the contact model and simulation in condition 2

In condition 3, with the consideration of the constraints considered in condition 1 and 2, introduce an assumption that the maximum static friction force is reached first on the nodes which in the lower right corner of slider. The assumption is not realistic which is drawn from the obtained results as shown in Fig.13.

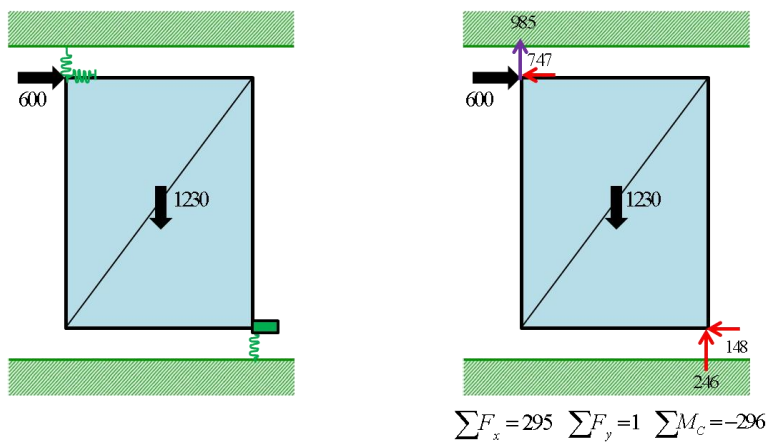

**Figure 13.** Diagram of the contact model and simulation in condition 3

In condition 4, with the consideration of the constraints considered in condition 1 and 2, introduce an assumption that the maximum static friction force is reached on the nodes which in the lower right and upper left corners simultaneously. The obtained results described in Fig.14 reveals that assumption is not realistic.

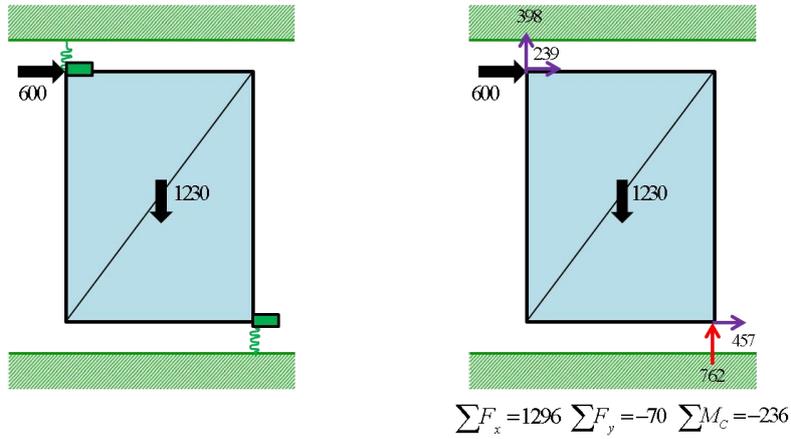

**Figure 14.** Diagram of the contact model and simulation in condition 4

In condition 5, with regards of the constraints considered in condition 1 and 2, a mechanical model is developed by introducing an assumption that the maximum static friction force is reached first on the nodes which in the upper left corner of slider. The solutions of the mathematical model reveal that six constrains are obeyed and confirm that the assumption is realistic, the mechanical model as illustrated in Fig.15. is truthful for describes the phenomenon during sticking.

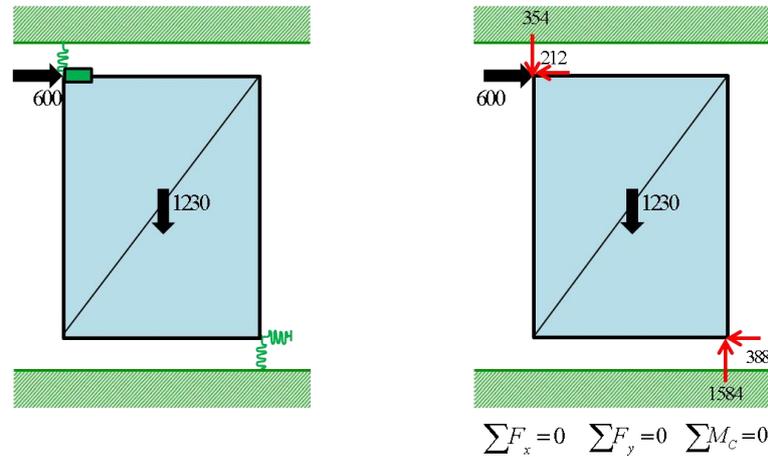

**Figure 15.** Diagram of the contact model and simulation in condition 5

Based on the equilibrium principles of force and moment, as described by equations (51) (52) and (53), a statics equations are formulated which are put in the form of a linear algebraic system. For the uniqueness of solutions of a linear algebraic system, inequality constraints (54) (55) and (56) can be imposed and obeyed only by the application of a trial-and-error method.

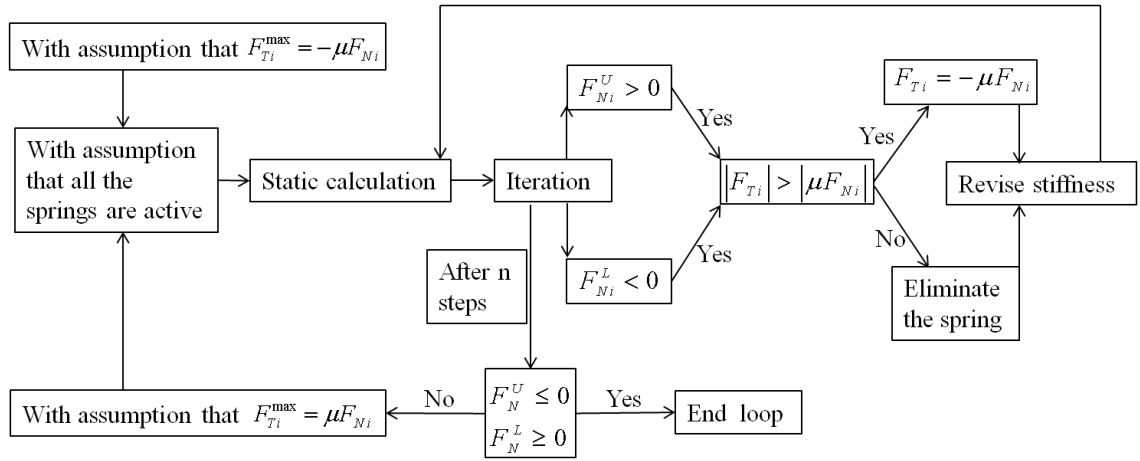

**Figure 16.** Block diagram of the trial-and-error method

### 3.3 The effect of small deformation on rigid body motion of translational joints

Based on the assumption that the influence of small deformation on rigid body motion is negligible[12], methods is presented for multibody system with frictional translational joints based on rigid body model[18,19]. This paper develops a method for a frictional translational joint based on deformable model. The obtained results reveal that it's Inappropriate to model contact problems with Coulomb friction and bilateral constraints based on the rigid body model, due to the assumption that neglecting the influence of small deformation on rigid body motion is improper sometimes. In some cases, just small deformation can affect the rigid body motion distinctly. The conclusion is significant for the simulation for the systems with frictional translational joints. The reason for the conclusion will appear below.

Contact at diagonally opposite surfaces can occur when the external force reached the corresponding value.

There are two types of model for contact region, based on the rigid body model, the contact model is described as point-to-surface as illustrated in Fig.17. With consideration of deformation, the point-to-surface contact model is transformed into surface-to-surface contact model as illustrated in Fig.18. Contact forces are pictured as red arrow and external forces are pictured as black arrow in Figs. 17 and 18.

In the surface-to-surface model, the normal contact forces acting on the slider are distributed forces. When the slider is considered as a rigid body, the contact model changed into point-to-surface and the normal contact forces are illustrated as concentrated forces. A comparison with moment arms of contact forces is made. As shown in Fig.17, the moment arm of contact forces about the mass centre of the rigid slider are larger than which described in Fig.18, which is the moment arm of contact forces based on flexible model. According to the equilibrium principle of moment, the absolute value of contact forces of deformable model is larger than that of rigid model. Hence the absolute value of tangential friction forces of deformable system is larger than that of rigid body system which is proportional to the normal contact forces during slipping. The differences of acting points of normal contact forces affect the tangential friction force and the dynamic motion in the tangential direction. Under uniform external forces, the absolute value of acceleration which is calculated by the flexible model is smaller than the absolute value of acceleration which is calculated by the rigid body model.

With the method mentioned in section 2.2.4, the influence of small deformation on rigid body motion can be calculated. The simulation is described in detail in the following subsection.

As the same as sticking, before proceeding, the contact region must be known so that the contact force can be calculated correctly, A trial-and-error method was used to solve the problem which is similar to the trial-and-error method mentioned in section 3.2.

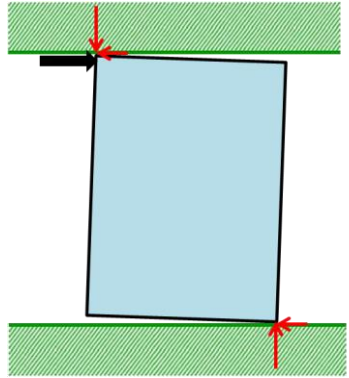 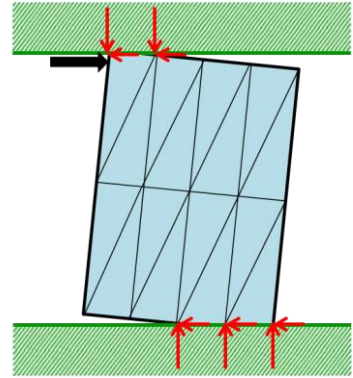

**Figure 17.** Diagram of contact forces based on rigid model    **Figure 18.** Diagram of contact forces based on flexible model

## 4  Numerical examples

Based on the method described in this paper, a specific code is developed using a FEM framework. Numerical simulation for translational joint with Coulomb friction and bilateral constraints which is calculated by the developed computer program is presented as follows.

A translational joint with Coulomb friction and bilateral constraints is presented as shown in Fig.19. The planar slider is made of an elastic isotropic, homogeneous material characterized by Young's modulus $E = 2.1 \times 10^{11}$ Pa and Poisson's ratio $o = 0.25$ (steel). The slider of mass $m$, length $a$, and width $b$, which is subjected to the action of the force $F^e$, moves in the guide for which the coefficients of kinetic friction and static friction are $\mu$ and $\mu'$, respectively. $C$ is the slider's center of mass, $A$ is the acting point of external force. The stiffness of normal fictitious springs and the tangential fictitious springs are calculated by equations in Penalty method. In this analysis 32 linear triangular elements, 27 nodes, and 54 degrees-of-freedom are used. Figure 19 shows the finite element mesh for the slider.

In all numerical computations presented here, the following values of parameters are taken:

$m = 125.6$ Kg, $\mu = 0.3$, $\mu' = 0.31$, $a = 2$, $b = 0.8$, $K_N = K_T = 1.05 \times 10^{11}$ N/m,

The initial conditions of the system are given as follows:

$q_0 = \dot{q}_0 = \ddot{q}_0 = 0$

The simulation is analyzed for four loading cases: (1) $F^e = 800N$, (2) $F^e = 1106N$, (3) $F^e = 1500N$.

Comparison is made between the results which are obtained by three methods: the simulation based on the method described in this paper, the calculation based on the rigid model and results using the FEM software ANSYS[22]. The value of penalty parameter $K^P_{ANSYS}$ equals to the default value in the ANSYS, $K^P_{ANSYS} = 2.1 \times 10^{11} \times 0.005$. Nodal contact forces and external forces are illustrated in the following Figs. Contact forces are pictured as red arrows, external forces and gravity force are pictured as black arrows. The geometry scale is changed in the Figures for representation. The resultant moment and resultant forces of the simulation are illustrated in the lower part of the Figures, the unit of forces and moments are $N$ and $N \cdot m$ respectively.

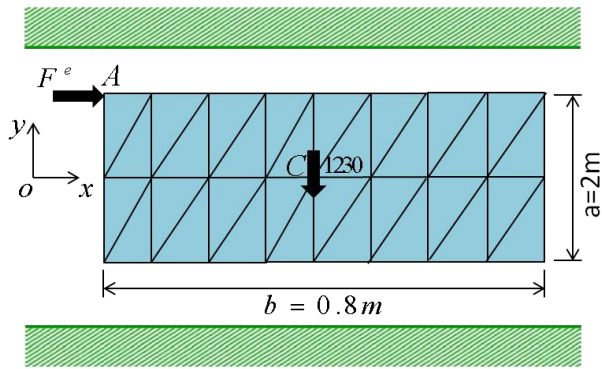

**Figure 19.** Diagram of the Numerical example

Case 1: $F^e = 800N$. The slider is in the condition of sticking and comes into contact with two surfaces. The calculation based on the rigid body model is obtained by introducing an assumption that the maximum static friction force is reached first in the upper left corner of slider, as shown in Fig.20. In Fig.21 the contact state is determined by the trial-and-error method mentioned in section 3.2. The analysis type of the simulation which is illustrated in Fig.22 is static, and analysis type of the simulation illustrated in Fig.23 is transient. Comparison between them reveals that the static analysis is more suitable than dynamic analysis, as to the simulation for system in the condition of sticking.

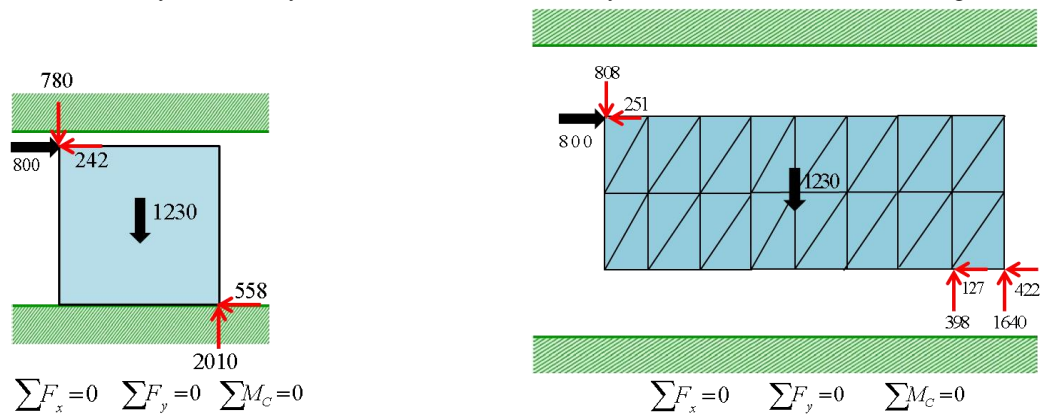

**Figure 20.** Closed-form solution based on the rigid body model

**Figure 21.** Simulation based on the model described in the paper

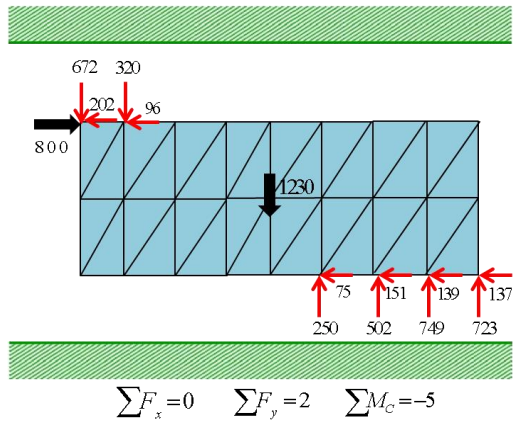
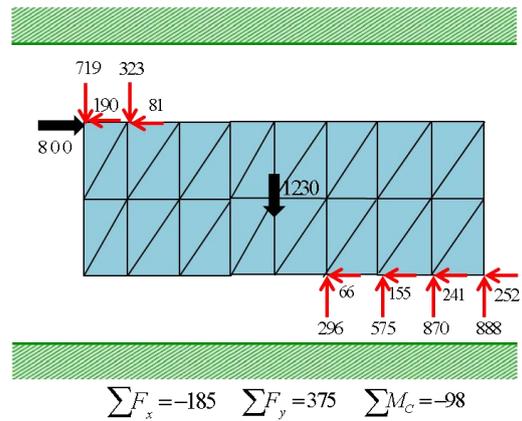

**Figure 22.** Result calculated by ANSYS with static analysis

**Figure 23**. Result calculated by ANSYS with transient analysis

Case 2: $F^e = 1106N$. The slider comes into contact with two surfaces and no sliding occurs. The calculation based on the rigid body model is obtained with the assumption that friction forces are equal to maximum static friction forces. Comparison is made with results obtained by different stiffness of normal and tangential fictitious springs. The results shown in Fig.25 is obtained with $K_N = K_T = 1.05 \times 10^{11}$ N/m which is calculated by the theory in Penalty method, and the results described in Fig.26 is obtained with $K_N = K_T = 2.1 \times 10^{11} \times 0.005$, which equals to the default penalty parameter in ANSYS. The comparison reveals the importance of the penalty parameter for the determination of contact region and contact forces.

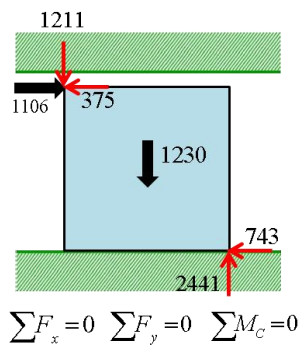
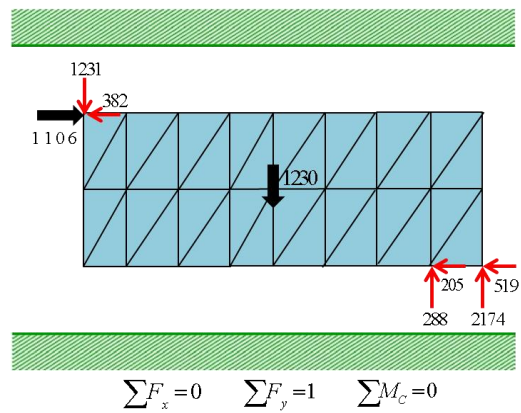

**Figure 24.** Result based on the rigid body model

**Figure 25.** Simulation based on the model described in the paper

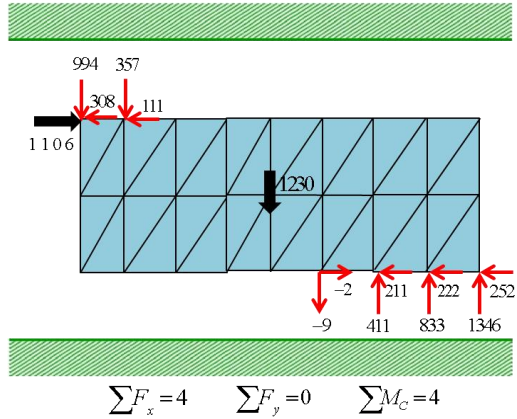
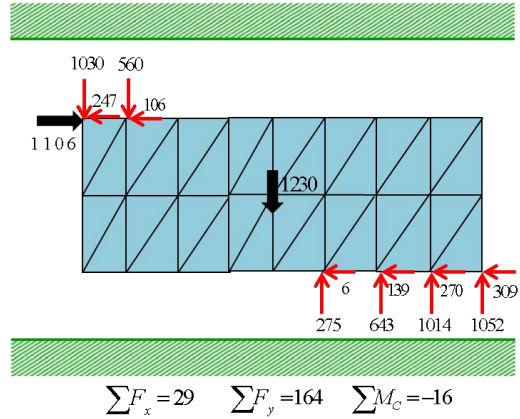

**Figure 26.** Simulation based on the model described in the paper

**Figure 27.** Result calculated by ANSYS

Case 3: $F^e = 1500N$. The slider comes into contact with two surfaces and is in the state of slipping.

The accelerations of slider are shown at right corner of Figs.28, 29, 30. The accelerations described in Fig.28 and Fig.29 reveals that small deformation affects the rigid body motion significantly. The equilibrium principle of moment and forces are not obeyed as illustrated in the Fig.30, shows the ANSYS is insufficient for 2D contact problems with Coulomb friction and bilateral constraints.

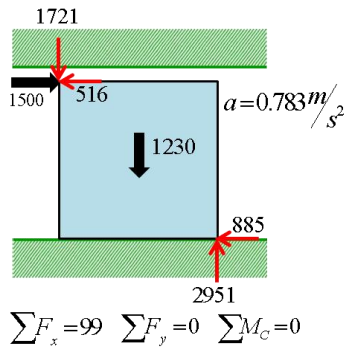
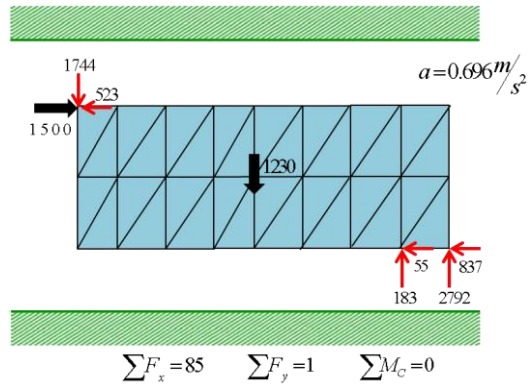

**Figure 28.** Closed-form solution based on the rigid body model

**Figure 29.** Simulation based on the model described in the paper

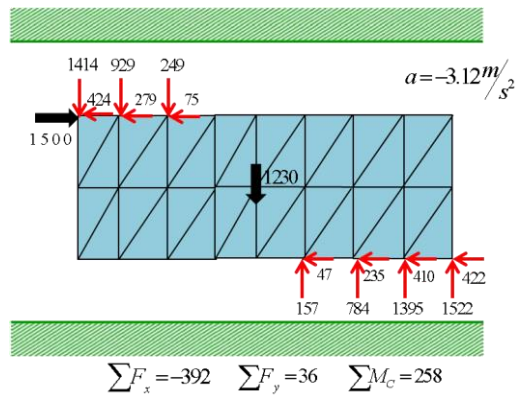

**Figure 30.** Result calculated by ANSYS

The numerical example illustrates the efficiency and capability of the method described in this paper for 2D contact problems with Coulomb friction and bilateral constraints, and two interesting phenomena can be extracted from the results.

## 5 Conclusion

A FEM method is developed based on the Plenty method which is able to analyse2D contact problems with bilateral constraints and Coulomb friction.

The simulation of the translational joint reveals two interesting phenomena: one is a special hyperstatic problem which is different from general hyperstatic problems, it can't be solved by general method by taking into account the deformation factor; the other is that small deformation affects the rigid motion significantly which is different from previous research.

**Acknowledgement**

The support of the National Natural Science Foundation of China (11072014) is gratefully acknowledged.